\documentclass{aa}
\usepackage{psfig}
\usepackage{graphics}
\begin{document}

   \thesaurus{06     
              (03.11.1;  
               16.06.1;  
               19.06.1;  
               19.37.1;  
               19.53.1;  
               19.63.1)} 
   \title{The Steep Spectrum Quasar PG1404$+$226 with ASCA, HST and ROSAT}

   \titlerunning{The Steep Spectrum Quasar PG1404$+$226}

   \author{M.-H. Ulrich
          \inst{1}
          \and
          A. Comastri\inst{2}
\and
S. Komossa\inst{3}
\and
P. Crane\inst{1,4}
          }

   \offprints{M.-H. Ulrich}

   \institute{European Southern Observatory, Karl-Schwarzschild-Strasse 2,
 D-85748 Garching, Germany\\
\and
Osservatorio Astronomico di Bologna, via Ranzani 1, I-40127 Bologna, Italy\\
\and
Max-Planck-Institute f\"{u}r extraterrestrische Physik,
Giessenbach Str.,  D-85748 Garching,
Germany\\
\and
Dept. of Physics and Astronomy, Wilder Laboratory, Dartmouth College,
Hanover N.H. 03755, USA
}   

   \date{Received xx, 1999; accepted xx, 1999}

   \maketitle

   \begin{abstract}
We present and discuss our observations of the Narrow Line quasar PG1404$+$226
($z_{em}$ = 0.098)  with ASCA and HST, and a re-analysis of our earlier
observations with ROSAT. The soft X-ray spectrum is very steep and displays
an absorption feature (edge or line at  $\sim$ 1.1 keV). We have applied
a variety of models to
the ASCA and ROSAT spectra without finding a completely satisfactory fit,
and the identification of the edge remains uncertain.
A satisfactory fit of the ASCA spectrum assuming that the edge is produced
by highly ionized iron (using the code absori in XSPEC)
is obtained with an overabundance of iron
by a factor $ \geq 25 $ compared to solar, a suggestion supported
by the extremely high equivalent width of the Fe K$\alpha$ line at 6.4 keV.
A warm absorber model
( based on Cloudy)
fitting the absorption feature with NeVII-NeX edges and assuming a peculiar
oxygen/neon abundance ratio is consistent with the ROSAT data but not
the ASCA data. Finally, it is also possible that the observed edge
is caused by  a OVIII or OVII edge or line,
blueshifted  by $z_{abs}$ = 0.2 to 0.5 depending on the
 specific identification, as has been suggested
previously for 2 other Narrow Line quasars, but there are no other features
in the UV and X-ray spectra in
support of this suggestion.

Two systems of UV absorption lines, one nearly
at rest in the source frame, the other blueshifted by $\sim 1900$ km s$^{-1}$
are identified in the HST/FOS spectra.
 Photoionization models indicate that
the UV absorption and the $\sim 1$ keV absorption are probably
caused by absorbers with different physical conditions.
PG1404$+$226 is one more case  of AGN where both UV and X-ray
absorption features
are detected, thereby increasing
further the significance of the previously noted statistical association of the
two types of absorbers.


   \end{abstract}

%

\section{Introduction}

Narrow Line Active Galactic Nuclei form
a distinct class of AGN on the basis of the properties of their
optical/UV spectrum: full width at half maximum (FWHM) of the hydrogen
lines and other lines in the range 500-2000 km s$^{-1}$,
intense high ionization lines, and intense FeII multiplets
(Osterbrock \& Pogge, \cite{osterbrock}; Shuder \& Osterbrock, \cite{shuder}).
The weakest of these AGN, the Narrow Line Seyfert 1 (NLS1),
and the somewhat brighter AGN with absolute optical luminosity
above but close
 to the lowest limit for quasars of $M_v$ = -23.4  have been
extensively studied in the X-ray range (e.g. Laor, Fiore, Elvis et al., 1994;
Boller, Brandt \& 
Fink, 1996;
Laor et al., 1997). 
Among all AGN, the Narrow Lines AGN tend to have the steepest
soft X-ray spectra (ROSAT), and some of them show
fast, large amplitude soft X-ray variability with occasionally
giant outbursts (e.g. Grupe 1996 and references therein; Boller et al 1996).
In the harder 2-10 keV band a comparative study of a large sample of
 NLS1 and broad line Seyfert 1s revealed that the 2-10 keV ASCA spectral
 slopes of NLS1 are significantly steeper than those of broad line
 AGN (Brandt, Mathur \& Elvis 1997).

 Recent BeppoSAX observations of a selected sample of bright NLS1
 (Comastri, Fiore, Guainazzi et al. 1998; Comastri, Brandt, Leighly et
al. 1998) over the broad 0.1-10 keV range
 indicate that a two component model provides an adequate description
 of the X--ray continuum. The relative strengths and slopes of the
 two components are different from those of broad line Seyfert 1s.
 The NLS1 are characterized by a stronger soft excess and, in general, have
 a steeper medium energy
  X-ray power law with respect to normal Seyfert 1s, but in PG 1404+226
the medium energy spectral index is not very different from that
of classical Seyfert 1s while its soft excess is strong, very steep and rapidly
variable like in other NLS1.
 The spectral behaviour of NLS1 suggests that the soft X-ray flux cannot be due
 only to disk reprocessing unless there is highly anisotropic emission.
 In the framework of the thermal models for the X-ray emission
 in Seyfert 1s (Haardt \& Maraschi 1993) a strong soft component could lead
 to a strong Compton cooling of the  hot corona electrons and to a steep
 hard tail. This hypothesis is also supported by the similarities
 between NLS1 spectra and those of Galactic black hole candidates in their
 high states first suggested by Pounds, Done and Osborne (1995) and
Czerny et al. (1996).

 The high states of Galactic black hole candidates are thought to be triggered
 by increases in the accretion rate possibly reaching
 values close to the Eddington limit.
 In this case the disk surface is expected to be highly ionized, in good
 agreement with the observation of a H-like Fe edge in TonS180
 (Comastri et al. 1998a).
 It should also be noted that a high $L/L_{Edd}$ ratio would also be consistent
 with the narrowness of the optical lines in NLS1 if the optical line
 producing region is virialized as suggested by Laor et al. (1997)

The quasar PG 1404+226 ($z$ = 0.098, V = 15) is one of the brightest members
of the class of Narrow Line AGN.
The optical spectrum of PG 1404+226 displays the characteristics of NLS1 with
FWHM (H$\beta$) $\simeq$ 830 km s$^{-1}$
and strong Fe II emission (Boroson \& Green 1992).\\
The observations with the ROSAT PSPC ($\sim$ 0.1--2.0 keV)
showed a very steep spectrum ($\Gamma \sim 3$) with
rapid flux (a factor 2 in 10 hours) and spectral variability typical of
NLS1 in  the X-ray range, and revealed a complex absorption feature
around 0.8-1.0 keV (Ulrich \& Molendi 1996, thereafter UM96). Time
resolved spectral analysis showed the data to be consistent with
a shift of  the absorption feature
to higher energy when the source brightens (UM96).

We have obtained a 40 ks ASCA observation of PG 1404+226  in order to
investigate the X-ray spectrum over a larger energy range and at
higher energy resolution than was possible with ROSAT. At the time of the
ASCA observations the absorption feature was around
1 keV, at an energy definitely higher than that of the  OVII and OVIII edges
at 0.74 and 0.87 keV commonly seen in ASCA and ROSAT
spectra of AGN (e.g. Reynolds 1997). Preliminary results can be found
in Comastri, Molendi \& Ulrich (1997, hereafter CMU97).
The ASCA and ROSAT data have been fit separately since
they have not been obtained at the same epoch.
The challenge presented by PG 1404+226 is the identification of the
absorption at $\sim 1.1$\ keV. Brandt et al (1994) found a similar feature
at 1.15 keV in Ark 564 and  considered several interpretations:
Neon edge, iron L edges, and outflowing material which would raise the
energy edge of OVIII (0.870 keV) to the observed energy.
They find none of them to be
satisfactory: neon edge because there is only a narrow range of
ionization parameter where it would be stronger than the OVIII edge; the
iron L edges would produce absorption at other somewhat higher energy
(1.358 keV); and the outflow seems unlikely because the source of its
kinetic energy
is unclear. Otani et al (1996) found a feature at $\sim 1$\ keV in
IRAS 13224-3809 for which they suggest an interpretation in terms of
outflow, or alternatively and considering that the the X-ray flux
of IRAS 13224-3809 varied by a factor 50 in two days ``the ionization state
of the medium is far
from equilibrium due to the violent variability''.
Krolik \& Kriss (1995) drew
attention to the fact that ``because the
ionization timescales of some ions may be as long as the variability
timescales in AGNs, the ionic abundances indicated by the transmission
spectra may not be well described by ionization equilibrium".
This point has recently
been re-investigated by Nicastro et al. (1999a) who pointed out also that
(i) Recombination times
can be longer than photoionization times resulting in
gaseous absorbers which are over-ionized with respect to the
equilibrium ionization state and (ii)  Collisional ionization
could be of comparable importance to photoionization.

 Hayashida (1997) found
an edge near 1 keV in H0707-495 for which he also
suggests an identification with
a blushifted OVIII absorption edge.
For PG 1404+226,  CMU97 suggested an overabundance of
iron, an interpretation which has the advantage to link the origin of the
1 keV absorption to the intense FeII lines present in the optical/UV
of this quasar and also of IRAS 13224-3809 and Ark 564.
The interpretation where the absorption near $\sim 1$\ keV
originates in hot gas outflowing at velocities in the range 0.2 - 0.5c
has also been proposed by Leighly et al (1997).

This paper presents an analysis of the ASCA observations,
a re-analysis of the ROSAT spectra with our warm absorber (WA) models
calculated with CLOUDY (Ferland 1993), and an analysis of HST spectra which we
have obtained in order to search for UV absorption/emission
lines. 

\section{Re-analysis of the ROSAT data}

The data reduction was carried out in a standard manner
and like in UM96 the data were split into ``high-state'
and ``low-state' depending on count rate.
The analysis was performed with our warm absorber models
based on CLOUDY.

\begin{table*}

\caption{\small X-ray spectral fits to the high-state (HS) 
and low-state (LS) {\sl ROSAT} data of PG 1404.
$\Gamma_{\rm x}$ was fixed to $1.9$. Errors in $U$, $N_{\rm WA}$
are about a factor 2--3.
}

\label{fiters}

      \begin{tabular}{ccccl}
      \hline
      \noalign{\smallskip}
        state & log $U$ & log $N_{\rm WA}$ &
                            $\chi^2_{\rm red}$ & model \\
      \noalign{\smallskip}
      \hline
      \noalign{\smallskip}
      HS/LS & & & 4.1/2.0 & single PL \\
     HS/LS & 1.0/0.7 & 23.7/23.4 & 2.5/0.9 & standard warm absorber \\
     HS/LS & 1.0/0.7 & 23.8/23.5 & 2.2/0.9 & emission+reflection added \\
     HS/LS & 0.7/0.3 & 23.3/22.9 & 1.3/0.9 & Ne = 4$\times$solar abundance \\
     HS/LS & 0.6/0.4 & ~~~23.1/23.1$^{(1)}$ & 1.2/1.0 & additional 0.1 keV soft
excess$^{(2)}$\\
      \noalign{\smallskip}
      \hline
  \end{tabular}

\noindent{\scriptsize $^{(1)}$ fixed to value derived for HS; 
$^{(2)}$ abundances reset to solar
}

\end{table*}

 The warm absorber is assumed to be of constant density,
of solar abundances (if not mentioned otherwise), and to be illuminated
by the continuum of the central point-like energy source. The spectral
energy distribution from the radio to the gamma-ray region
consists of piecewise powerlaws with, in particular, an energy index
in the EUV, $\alpha$$_{uv,x}$, of 1.4 and an X-ray photon index $\Gamma$$_{x}$
of 1.9. A black-body-like soft excess is added in some models.
The column density N$_{WA}$ of the warm material (i.e. the total column density
in Hydrogen) and the ionization parameter U
are determined from X-ray spectral fits (Table 1).
 U is a measure of the number rate of ionizing photons above the Lyman limit
and is defined by
\begin {equation}
U=Q/(4\pi{r}^{2}n_{\rm H}c)
\end {equation}

where $Q$ is the number rate of incident photons above the Lyman limit,
$r$ is the distance between
central source and absorber, $c$ is the speed of light, and
$n_{\rm H}$ is the hydrogen density
(fixed to 10$^{9.5}$ cm$^{-3}$ unless noted otherwise). As detailed below,
several models give acceptable fits to the low-state spectrum but
none gives a really good fit to the high-state spectrum (which has a
higher signal/noise ratio).

\subsection{Results with a simple power law}

A simple power law plus cold absorption gives a very poor fit.
The cold absorbing column generally tends to slightly
underpredict the Galactic value. This may reflect
some low energy
calibration uncertainties, or the presence of an additional
spectral component in the form of a very soft excess.
Adding a soft excess parameterized as a black body
which contributes only at very low energies (e.g. in NGC4051,
Komossa \& Fink 1997; Ton S180, Fink et al. 1997)
improves the fit which, however, remains unsatisfactory.

\begin{figure}
\psfig{file=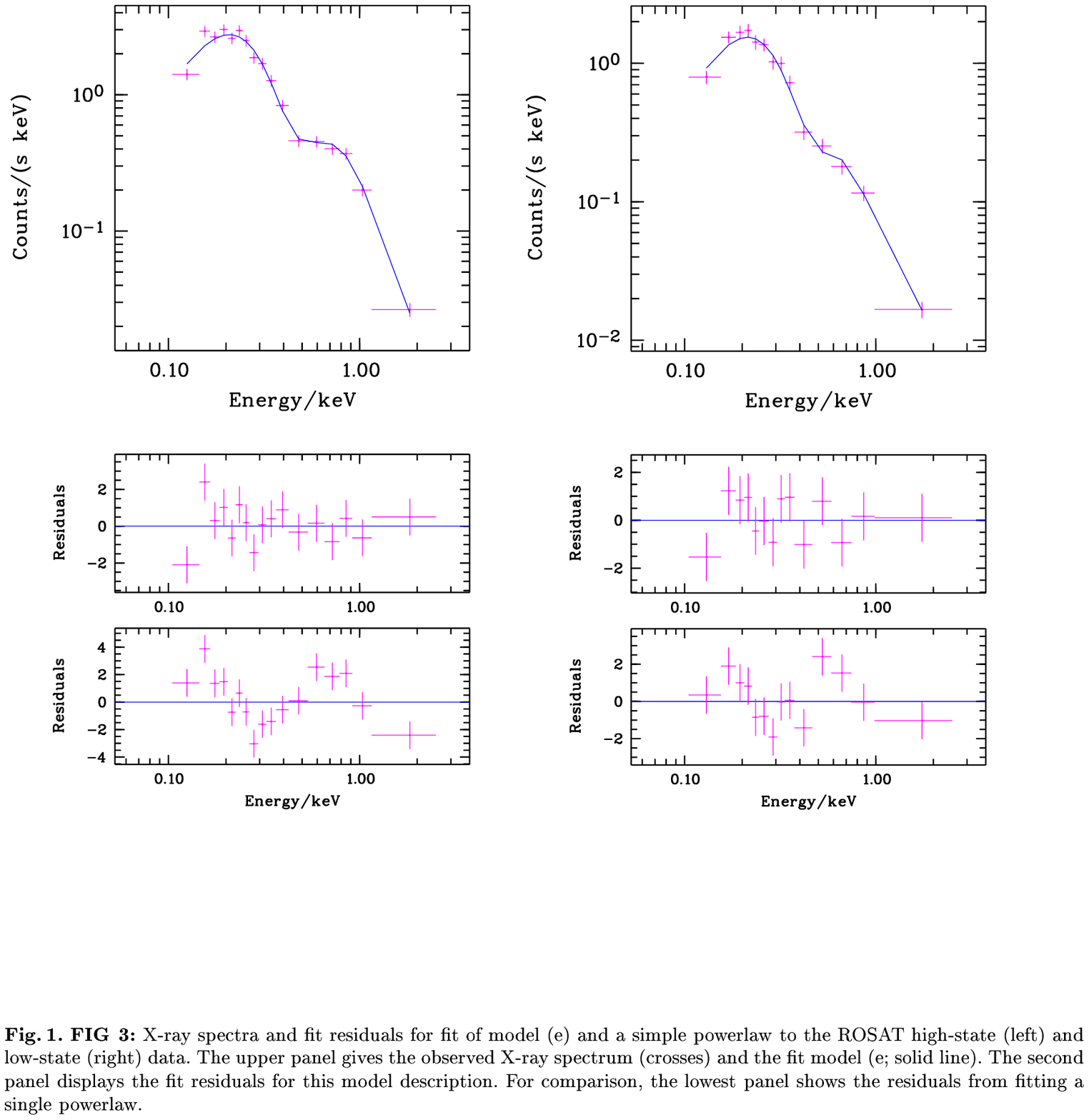,width=\hsize,clip=}
\caption{X-ray spectrum (crosses) for fit of model (e) (continuous line) 
to the ROSAT
high state (left) and low state (right). Immediately underneath are
the corresponding residuals for this model. The bottom panels are the
residuals for a single power law fit of the ROSAT high state (left)
and low state (right).}
\label{}
\end{figure}

\subsection{Results of the Spectral fits with Warm absorber models}

We have examined a family of models (labelled a,b,..e; Table 1)
which include a  warm absorber plus
a power law component with a photon index
$\Gamma$$_{x}$ fixed at 1.9.

\subsubsection{Model a and b: solar abundances}

{\it  Model a:}
The inclusion of a `standard' warm absorber model
clearly improves the fit but does not  yield an
acceptable $\chi^2$, with systematic residuals
    remaining near 1 keV. The reason is that, even for very high
    ionization parameters, the model has a strong oxygen absorption
at 0.87 keV, even
    stronger than the neon absorptions (NeVII at $\approx$ 1.1 keV and
    NeX at $\approx$ 1.36 keV).  This is
    exacerbated at high-state because
    the high-state data require absorption around the location of
    the neon edges to dominate. (The same results hold for a steeper
    underlying powerlaw.) \\
{\it  Model b:}     The addition of the emission and reflection components
    of the warm material, calculated for a covering
factor of 0.5  results in
    a slightly improved fit, but the high-state data are still not
    well matched.

\subsubsection{Model c: non-solar abundances}

    One way to make the neon absorption dominate over the oxygen
    absorption is to introduce deviations from solar abundances, with
     overabundant neon, or underabundant oxygen.\\
    Several deviation factors were studied between an oxygen abundance of
    up to O = 0.2 x solar and a neon abundance of  Ne = 4 x solar.
    These models strongly improve the quality of the fit
    and the values of $\chi^2$ reach acceptable values (for the ROSAT spectra).
    The best fit has
an overabundance of neon of $\sim 4$ times the solar value.\\
    We note that  while
non solar abundances have been reported in a number of AGN/quasars
(e.g. Hamann \& Ferland 1993, Netzer 1997) the ratio O/Ne is expected to remain
 close to its solar value in all known astrophysical situations (On the
other hand, Netzer occasionally depletes {\it oxygen only} in his
photoionisation
models, e.g. Marshall et al. 1993).
In any event, the model with overabundance
of neon is not compatible with the  ASCA data (Section 3.2.2).

\subsubsection{Model d and e: additional soft excess (and solar abundances)}

{\it Model d:}
    Motivated by the ASCA evidence for the presence of a soft
    excess, a sequence of models was calculated with an additional hot
    BB component of kT = 0.1 keV. This component was included
    in the ionizing spectral energy distribution that illuminates
    the warm absorber, i.e. the change in
    ionization structure of the warm material was self-consistently
    calculated.
    A successful description of both, high- and low-state data
    is possible with solar abundances. Note that two-component models
{\it without edge}, specifically, power law + blackbody and
power law + bremstrahlung
were found to provide satisfactory fits of the ROSAT data at low
state. For the high state data the  power law + blackbody model gives
$\chi^2$ = 67/20 (UM96).\\


 {\it Model e:}
    U and N$_{WA}$ have above been treated as free
    parameters, and the fits tend to give slightly
    different column densities in low- and high-state data, whereas
    this quantity is not expected to vary within short time-scales (10
hours or less).
    We therefore have searched for the best fit model (model e)
 to the high- and low-state
    data in which the column density is identical in both states and only the
    ionization parameter U and the strength of the BB component
    are allowed to vary between states.
    We find a successful model with log(N$_{WA}$) = 23.1, log(U$_{high}$) = 0.6,
log(U$_{low}$) = 0.4, and a contribution of the blackbody to the ionization
    parameter of log(U$_{BB}$) = log(U$_{PL}$) - 0.9 in the high-state data
and log(U$_{BB}$) = log(U$_{PL}$) -1.5 in the low-state data (Fig. 1;
U$_{BB}$ and U$_{PL}$ are the ionization parameters
  related to the blackbody component, and powerlaw component, respectively.
U is a measure of the number rate of ionizing photons above the Lyman
  limit. See Eqn. (1) for the definition of U).

\section{ASCA Observations and Spectral Analysis}

\subsection{Observations}

PG 1404+226 was observed by ASCA (Tanaka, Inoue \& Holt 1994)
July 13-14, 1994 with the
Gas Imaging Spectrometers (GIS2/GIS3) for a total effective exposure time
of 35000 s, and with the Solid--state Imaging Spectrometers (SIS0/SIS1) for
about 29000 s.

\begin{table*}
\caption{ASCA spectral analysis}
\label{}

\begin{tabular}{lccccc}

\hline
\noalign{\smallskip}

Model & 
$\Gamma$ or $kT$ $^{(1)}$ &
$\Gamma$ $^{(2)}$ &
$E_{edge}$/$E_{cut}$ $^{(3)}$ &
$\tau$/$E_{fold}$ $^{(4)}$ &
$\chi^{2}$/dof \\

\noalign{\smallskip}
\hline
\noalign{\smallskip}

PL            & $3.48 \pm 0.10$	& ... &	... & ... & 438.3/146 \\

PL+PL         & $3.95 \pm 0.15$ & $0.85^{+0.21}_{-0.31}$ & ... & ... & 
291.5/144\\

BR+PL 	      & $246^{+17}_{-15}$ & $1.48 \pm 0.18$ & ... & ...	&  235.7/144 \\

BB+PL	      & 126 $\pm$ 4 & $1.62^{+0.15}_{-0.11}$ & ... & ... & 196.0/144 \\   

BB+PL+EDGE    & $138\pm 6$ & $1.92^{+0.19}_{-0.16}$ & $1.07 \pm 0.03$ & 
$0.88\pm 0.26$& 158.6/142 \\ 

CUTBB+PL   & $139\pm 8$ & $1.91^{+0.10}_{-0.15}$ & $0.94^{+0.05}_{-0.07}$ 
& 51$^{+96}_{-45}$ & 159.5/142 \\

\noalign{\smallskip}
\hline
\end{tabular}
\smallskip

\noindent{\scriptsize 
$^{(1)}$ Power law slope or thermal model temperature (in eV) for the
soft component; \\
$^{(2)}$ Power law photon index for the hard component; \\
$^{(3)}$ Edge energy / Exponential cutoff energy (in keV); \\
$^{(4)}$ Edge optical depth / e-folding energy of the exponential 
cutoff in eV
}

\end{table*}

\begin{figure}
\psfig{file=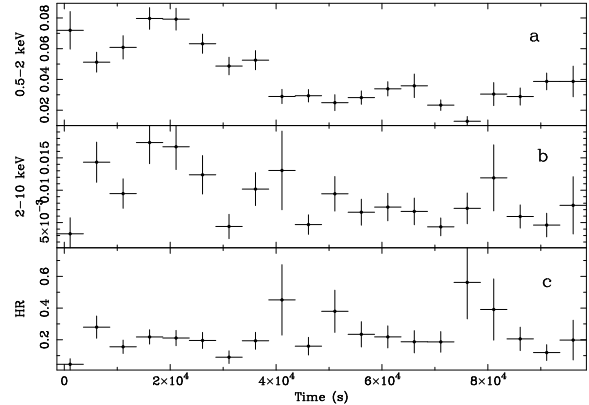,width=\hsize,angle=270}
\caption{
Intensity and hardness ratio light curves of PG 1404+226
        during the ASCA observations (for ROSAT observations see fig 3 in
        UM96). Panel a) SIS0 light curve (counts/s) in the 0.5-2 keV
        energy range.
        Panel b) SIS0 light curve (counts/s) in the 2-10 keV energy range.
        Panel c) Hardness ratio (0.5-2/2-10 keV) light curve.}
\label{}
\end{figure}

The SIS was operating in 1-CCD mode and all the
data were collected in Faint mode. Standard criteria for the good
selection intervals have been applied (i.e. Cut-off rigidity $>$ 7 for GIS
and $>$ 6 for SIS, minimum elevation angle from the earth $>$ 5 degrees and
minimum bright earth angle $>$ 25 degrees for SIS) as well as DFE and echo
corrections.
The background subtracted count rates for PG 1404+226 are
0.049,0.038,0.014,0.017 c/s in S0, S1, S2, S3 respectively.

Large flux variations have been detected.
In panels a and b of Fig. 2
 we show the soft (0.5--2.0) and hard
(2.0--10 keV) SIS0 light curves.
The SIS0 light curve in the soft band (Fig. 2a ) shows a 
factor $\sim$ 4 of amplitude variability
with a doubling timescale of  $\sim$ 8000 s.
The variability pattern in the two bands is well correlated suggesting
that the hard and soft fluxes varied in the same way.

A detailed spectral resolved temporal analysis
is hampered by the low counting statistics especially at high ($>$ 2 keV)
energies.
In order to have some indications on the spectral variability
we have performed a hardness ratio analysis for the high (first part of the
observation) and the low (second part of the observation)
state shown in Figure 2.
The hardness ratio (Fig. 2c)
has been defined as HR = (H-S)/(H+S) where H and S
are the counts in the 2--10 and 0.5--2 keV bands respectively. The results
for the high and low state are $HR_{High state} = -0.79 \pm 0.06$ and
$HR_{Low state} = -0.75 \pm 0.04$ suggesting that the spectrum is rather soft
in both states. The 2--10 keV flux is rather
  low and as a consequence both the 2-10 keV light curve and the hardness
  ratio light curve are noisy with no clear evidence of
  spectral variability  in contrast
to the spectral change during the ROSAT observations.
The spectral analysis has thus been performed on the spectrum
accumulated over the entire duration of the ASCA observation.
The average luminosity is comparable with the ROSAT low state but the
absorption  feature is at high energy as during the ROSAT high state.\\

\subsection{Spectral Analysis}

GIS and SIS spectra were binned with more than 20 counts/bin
in the 0.7--10 keV and 0.5--10 keV energy ranges respectively.
Since the spectral parameters obtained by fitting the four detectors
separately were all consistent within the errors, the data were
fitted simultaneously to the same model.

Given the very low Galactic column density toward PG 1404+226
($2 \times 10^{20}$ cm$^{-2}$; Elvis, Lockman \& Wilkes 1989)
and the low sensitivity
of ASCA detectors for small column densities, all the spectral fits
have been performed with $N_H \equiv N_{HGal}$. This choice is
also corroborated by the ROSAT results (UM96).

A single power law model clearly does not provide an acceptable fit (Table 2)
The power law fits reported in Table 2 suggest that at least
two components are required to fit the overall spectrum.
We note that the slope of the hard power law,
which could not have been detected by ROSAT, is consistent with the average
quasar slope (Comastri et al. 1992)
while the slope below 2 keV is consistent with the UM96 findings.

In order to model the broad band (0.4--10 keV) spectrum
we tried 4 types of double component fits (See Table 2 for details):
a double power law, bremsstrahlung plus power law, blackbody
plus power law, and cut-off blackbody plus power law (the cut-off is of
the form exp~$[-(E-E_c)/E_f]$ for energies larger than the cutoff energy
$E_c \sim 0.94$ keV. The depth of the cut-off is related to $E_f$,
a small value correponding to a very steep decline of the intensity; here
$E_f \sim 0.05$ keV).

The first two models give unacceptable fits. The only two
acceptable descriptions of the data are
the cut-off blackbody plus power law (with a cut-off at 0.94
keV) and the blackbody plus power law plus
absorption edge model around 1 keV .
The addition of the absorption edge to the simple blackbody plus
power law model produces an improvement in the fit which is
significant at $> 99.99\%$ (F-test) and the residuals are featureless
in agreement with the results of Leighly et al (1997).

\subsubsection{Is iron overabundant?}

As absorption edges are the signature of warm absorbing material, we
fit the
warm absorber model available in XSPEC 9.0
(the model is `absori' and the
iron abundance is a free parameter; it was developed by P. Magdziarz \&
A. Zdziarski following Done et al 1992 and Zdziarski, Ghisellini, George 
et al. 1990).
Given the relatively large numbers of free parameters in this model
we have fixed the temperature of the warm material at $T = 10^{5}$ K
(see for example Reynolds \& Fabian 1995; the fits are not sensitive to
the temperature, a change of log{\it T} from 4.5 to 6
would make no significant difference) and the iron abundance at the
solar value. 
In addition we have considered only the 0.5--3.0 keV energy range in order to
avoid contamination from the high energy component.
We were not able to find any acceptable solution 
as can be judged from figure 3.
Leaving the iron abundance free to vary the improvement is significant 
at $> 99.9$\%  (F-test) and the residuals featureless (Fig 4).
The resulting parameters are:
$\xi = 3000_{-1000}^{+500}$ erg cm s$^{-1}$,
$N_H = 8_{-1}^{+9} \times 10^{20}$ cm$^{-2}$, Iron abundance $>$
25 solar and power law photon slope of $3.55 \pm 0.12$.
(All the quoted errors are at 90\% confidence for one interesting
parameter, $\chi^2_{min}$ + 2.7).

\begin{figure}
\psfig{file=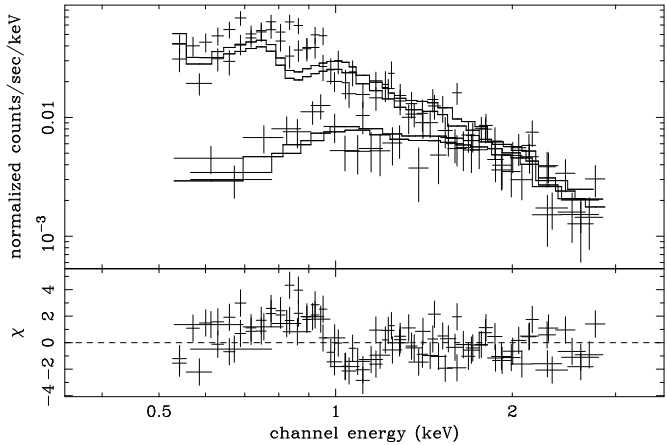,width=\hsize,angle=270}
\caption{The 0.5--3 keV ASCA spectrum fitted with a power law plus
  a warm absorber model with the iron abundances fixed at the solar
  value.
  The four ASCA detectors are fitted simultaneously leaving the
  relative normalizations free to vary.}
\label{}
\end{figure}

\begin{figure}
\psfig{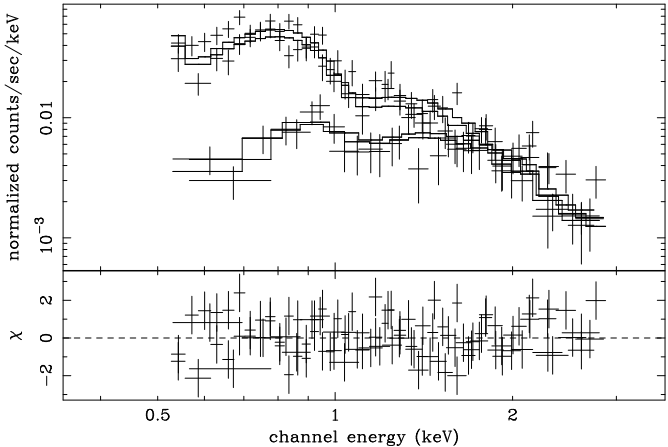}

\caption{
The 0.5--3 keV ASCA spectrum fitted with a power law plus
  a warm absorber with super-solar iron abundances.
  The four ASCA detectors are fitted simultaneously leaving the
  relative normalizations free to vary.}
\label{}
\end{figure}

%

\begin{figure}
\psfig{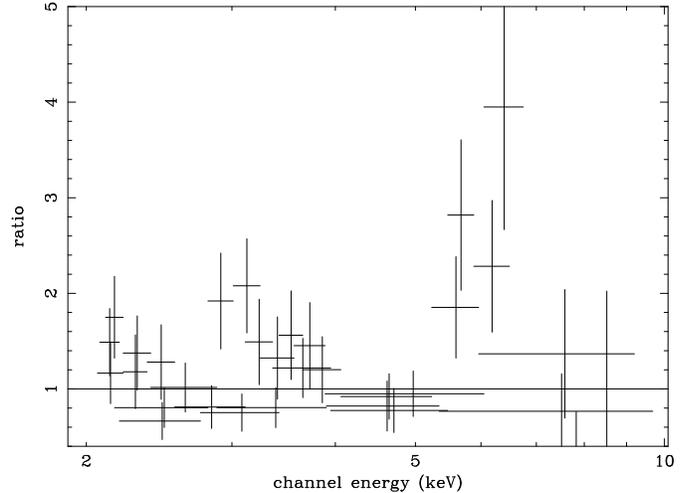}
\caption{The residuals of a power law fit to the ASCA data in the 
2--10 keV energy range. The two GIS and the two SIS have been added.
A line--like excess is evident around 6 keV (observed frame).}
\label{}
\end{figure}

The shape of residuals around 6--7 keV are suggestive of iron line emission
(Fig 5).
With the addition of a narrow line the improvement in the fit is, however,
significant only at the 2 $\sigma$ level. The derived parameters are
$E_{K\alpha} = 6.5 \pm 0.3$ keV and EW = $1290 \pm 680$ eV.

It is interesting to note that the huge equivalent width of the
iron line is qualitatively
 consistent (see figure 3 in Reynolds, Fabian \&  Inoue, 1995)
 with the supersolar iron abundance found by
fitting the data with the warm absorber model.

\subsubsection{Non-solar abundances of Neon or Oxygen?}

We have also attempted to fit some warm absorber models
(cloudy based, and with blackbody + power law continuum)
  to the ASCA spectrum and have found no satisfactory fit - the reason
being that the
edge-like feature at 1 keV is too deep and too narrow for these models.

\subsubsection{Resonant Absorption ?}

It has been recently shown that the spectra emerging from ionized gas
are strongly modified by resonant absorption if the dispersion velocity
in the gas is of the order of 100 km s$^{-1}$ or greater
(Nicastro, Fiore, Matt 1999b). The X--ray warm absorber features
are also strongly dependent from the shape of the ionizing continuum.
For soft X--ray spectra as steep as the one observed for PG 1404+226
several resonance absorption lines from Fe L, Mg, Si and S are predicted
between 1 and 2 keV. Such blend of lines would appear as negative residuals
in low resolution ASCA spectra.
The $\sim$ 1 keV absorption feature of PG 1404+226 could be, at least in part,
accounted for by resonant absorption. Signatures of this process have been
looked for in the HST UV spectrum (see below).

\subsubsection{Is the $\sim 1$ keV absorption the blueshifted OVIII edge?}

Finally, we have also considered the possibility that the absorption
at 1.07 keV is the blueshifted OVIII edge whose rest frame energy is 0.87 keV.

We have calculated
the optical depth expected for the Neon edges at rest energies 1.1
and 1.36 keV which would be blueshifted to 1.29 and
  1.55 keV respectively . For the 1.1 keV line
the instrumental  upper limit on
$\tau$ is 0.14 (90\% confidence level) and is 0.27 for the 1.36 keV edge.

We note that with  a `standard' warm absorber in relativistic outflow
the OVIII edge is never the ``only'' edge in the spectrum. If the
absorber is lowly ionized, there is OVII co-existing and if it is very highly
ionized, OVII becomes negligible, but one cannnot avoid having some Neon
absorption. Although the exact ratios of the optical depths $\tau$
depend on the input
parameters, we thought it would be useful to check whether we could detect any
other edge, or check how strict the upper limits are. We take NGC 4051
as basis for
comparison. In NGC 4051, the absorber is  rather highly ionized,
with OVIII being the strongest edge with $\tau$=
$1.1 \pm 0.4$, then NeX with $\tau$=$0.8 \pm 0.4$.
The OVII edge is weaker with $\tau$=0.35; Komossa \& Fink, 1997.
Using the ratio $\tau$$_{OVIII}$/$\tau$$_{NeX}$
 as typical for the relative depths in
a highly ionized absorber, we can conclude that our non-detection
of edges other than the one near 1.1 keV in PG1404+226
is still consistent with a standard warm absorber
i.e. our upper limits on $\tau$ are not strict enough
to rule out a warm absorber in relativistic outflow.

\section{HST and IUE Observations}

\subsection{Observations}

Four spectra were taken in February 1996 with HST/FOS and
gratings G130, G190H, G270 and G400 with integration times 2300,
530, 120 and 120 seconds respectively. The total observed wavelength range
covered
is 1087 - 4773 {\AA} with a nominal resolution of 1300 (Fig.6).
The spectra were taken through the 0.86 arcsec diameter aperture. Standard
reduction procedures were performed. The wavelength scale of the spectrum taken
with G190H was shifted by +1.5 {\AA}, to be consistent with
 the other spectra.


\begin{figure}
\psfig{file=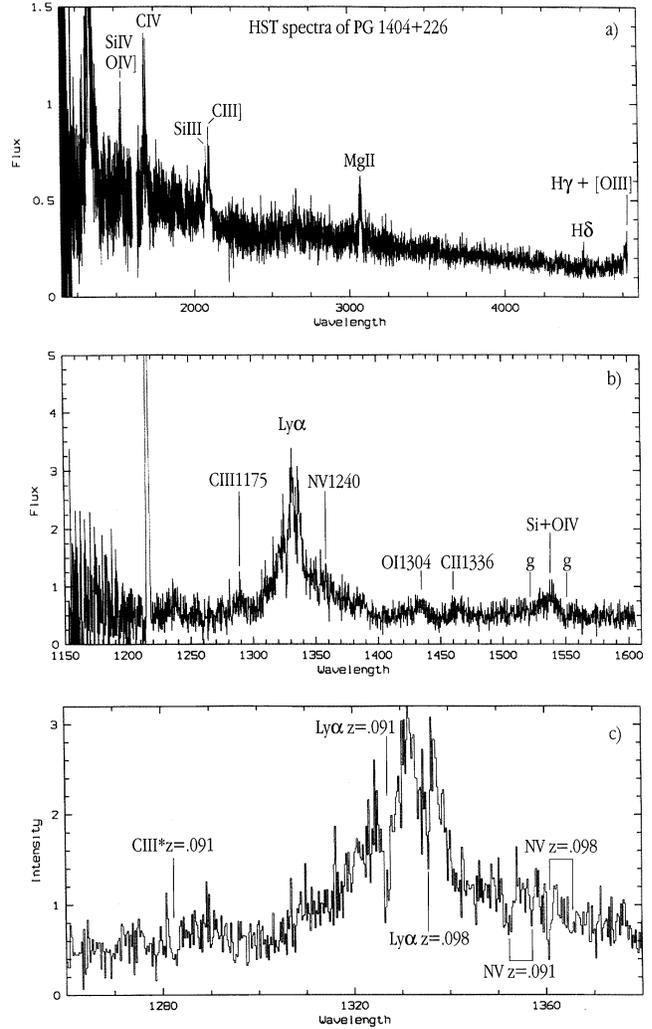,width=\hsize,clip=}

\caption{HST spectra of  PG1404+226: a) in the  observed
range 1150-4773 \AA\.  b) the FOS/G130 spectrum and the emission lines
identification c) the Ly$\alpha$  and NV absorption lines
identification.}
\label{}
\end{figure}

PG 1404+226 was observed with IUE on July 1994, 2 days after the ROSAT
observations. The spectrum, SWP 51419, has a total integration time
of 315 minutes (accumulated in 12 parts), and was taken through the large
aperture and at low dispersion. Compared with the HST spectra taken
18 monthts later the continuum flux in July 1994 is $\approx$ 1.3
times brighter in the
common observed wavelength range 1265 to 1400{\AA} but the Ly$\alpha$ line
 kept the same intensity. The modest S/N and spectral resolution of the
IUE spectrum would prevent
the detection of  the absorption lines seen in the HST
spectra. No change in the emission/absorption profile of Ly$\alpha$ (Fig.7)
can be detected by comparing the IUE spectrum and the FOS-G130 spectrum
rebinned at 2{\AA}.


\begin{figure}
\psfig{file=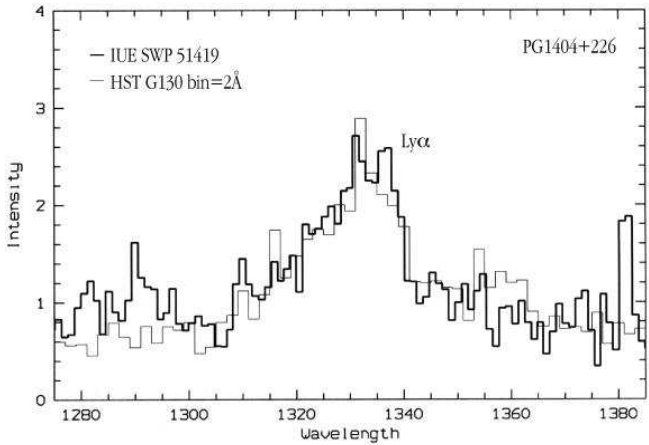,width=\hsize,clip=}

\caption{The Ly$\alpha$ line in the IUE spectrum of 16 July 1994 (thick line)
and the HST spectrum of February 1996 rebinned with 2 A bins (thin line).}
\label{}
\end{figure}

\subsection{Continuum and emission lines}

Table 3  lists the emission line intensities (H$_{0}$ =
50 km s$^{-1}$ Mpc$^{-1}$, q$_{0}$ = 0, distance = 617 Mpc).
We note the presence of some weak emission lines:
(1) an unidentified line at 1175{\AA} (rest wavelength 1070{\AA})
noticed in a few other quasar spectra (Laor et al 1995; Hamann et al 1997)
(2) a line at 1290 \AA\ which we identifiy with CIII$^*$1175.7. 


\begin{table}

\caption{Emission Lines Intensity in the HST Spectra of PG 1404+226}

\label{}

\begin{tabular}{llr@{}lr@{}l}

\hline 
\noalign{\smallskip}

 & & \multicolumn{2}{c}{Observed Flux} 
     & \multicolumn{2}{c}{Intrinsic Flux} \\
 & & \multicolumn{2}{c}{in 10$^{-14}$} 
     & \multicolumn{2}{c}{in 10$^{41}$} \\
Identification & $\lambda_{\rm meas}$ 
     & \multicolumn{2}{c}{ergs s$^{-1}$ cm$^{-2}$} 
       & \multicolumn{2}{c}{ergs s$^{-1}$} \\

\noalign{\smallskip}
\hline 
\noalign{\smallskip}

OVI 1033.83        & 1135.15 &    t&             &   t& \\
HeII 1085.15       & 1191.5  &    1&.2           &   5&.4 \\
CIII 1175.70       & 1290.0  &    1&.2           &   5&.4 \\
Ly$\alpha$ 1215.67 & 1334.8  &   54&             & 245& \\
NV 1240.15         & 1361.7  &   --&             &  --& \\
OI 1304.46         & 1432.25 &    2&.5           &  11&.3 \\
CII 1334.53        & 1465.0  &    1&.2           &   5&.4 \\
SiIV 1396.76       & 1533.6  &    3&.5           &  15&.9 \\
NIV] 1486.50       & 1632.2  &   --&             &  --& \\
CIV 1549.05        & 1700.85 &   11&.5           &  52&.2 \\
HeII 1640.50       & 1801.3  &   --&             &  --& \\
SiIII] 1892.03     & 2074.75 & SiIII]+&CIII] 9.9 &  45&.0 \\
CIII] 1908.73      & 2093.9  & \\[3mm]

MgII 2795.53       & 3069.0  &    4&.5           &  20&.4 \\
MgII 2802.70       & 3078.1  &   --&             &  --& \\[3mm]

H$\delta$          & 4945.4  &    1&.0           &   4&.5 \\
H$\gamma$ +[OIII]  & ~~~--   &    1&.3           &   5&.9 \\

\noalign{\smallskip}
\hline
\noalign{\smallskip}

\multicolumn{6}{l}{{\bf Footnote:} 
	H$_0$ = 50 km s$^{-1}$ Mpc$^{-1}$, q$_0$ = 0} \\

\end{tabular}

\end{table}

This line is
seen in HST spectra of IZw1 and Laor et al (1997) suggest that it
 is produced by resonance
 scattering of continuum
photons by CIII$^*$ ions, a mechanism which requires large velocity gradients
($\approx$ 1000 km s$^{-1}$) within {\it each} emitting cloud of the BLR.

\subsection{The two systems of absorption lines and their likely origin}

We identify two absorption systems in the HST spectra
which, in the source frame, are separated by $\sim 1920$ km s$^{-1}$.
In the `blue' system the absorption lines appear in the blue
flank of the Ly$\alpha$ and CIV emission lines at 800 km s$^{-1}$
from the peak. In the `red' system the absorption lines appear on the red
flank of the emission lines at 1100 km s$^{-1}$ from the peak.
It is known that in radio quiet AGN/Quasars, 
the high ionization lines such as CIV
are blueshifted with respect to the
systemic velocity by a few hundred to a few thousand km s$^{-1}$
(e.g. van Groningen 1987, Corbin 1995, Sulentic et al. 1995)
this blueshift being generally
interpreted as evidence for a wind outflowing from the face of the
accretion disk turned towards us. On this
basis, we argue that the red absorption system is close to zero
velocity in the source frame while the blue system originates from
material which has an outflow velocity of $\sim 1900$ km s$^{-1}$ (towards us
with respect to the quasar).


\begin{table}
\caption{Absorption Lines in the HST Spectra of PG 1404+226}
\label{}

\begin{tabular}{llll}

\hline 
\noalign{\smallskip}

$\lambda$ (\AA) & Identification & $z$ meas. & EW (\AA) \\
meas.           &                &           & meas. \\

\noalign{\smallskip}
\hline 
\noalign{\smallskip}

\multicolumn{4}{l}{Galactic lines:} \\[3mm]

1206.5  & SiIII 1206.50      &          & 0.6 \\
1259.8  & SiII 1260.42       &          & ~~t $^{(1)}$ \\
2795.5  & MgII 2795.53       &          & 0.8 \\
2802.5  & MgII 2802.70       &          & 0.95 \\
2851.7  & MgI 2852.13        &          & ~~t \\

\noalign{\smallskip}
\hline 
\noalign{\smallskip}

\multicolumn{4}{l}{Abs. System at z=0.091} \\[3mm]

1282.5  & CIII*1175.70       & 0.09084  & 0.6 \\[3mm]

1327.0  & Ly$\alpha$ 1215.57 & 0.091579 & 1.2 $^{(2)}$ \\
1352.4  & NV 1238.81         & 0.091700 & 0.6 \\
1356.3  & NV 1242.79         & 0.091330 & 0.2 \\[3mm]

1521.2  & SiIV 1393.76       & 0.09144  & 0.3 \\
        & SiIV 1402.77       &          & ~~-- $^{(1)}$ \\[3mm]

1688.5  & CIV 1548.19        & 0.090630 & 1.1 $^{(2)}$ \\
1691.5  & CIV 1550.76        & 0.090750 & 0.8 \\

\noalign{\smallskip}
\hline 
\noalign{\smallskip}

\multicolumn{4}{l}{Abs. System at z=0.098} \\[3mm]

1334.60 & Ly$\alpha$ 1215.57 & 0.097830 & 0.95 $^{(3)}$ \\[3mm]

1360.65 & NV 1238.81         & 0.098360 & 0.4 \\
1365.45 & NV 1242.79         & 0.098690 & 0.25 \\[3mm]

  --    & SiIV 1393.76       &          & ~~-- \\
1538.5  & SiIV 1402.77       & 0.0967   & ~~t \\[3mm]

1699.1  & CIV 1548.19        & 0.097480 & 0.55 \\
  --    & CIV 1550.76        &          & ~~-- \\[3mm]

2860.0  & FeII2599.4 ?       & 0.10     & 0.7 \\

\noalign{\smallskip}
\hline 

\end{tabular}
\smallskip

\noindent
{\bf Footnotes:}

\noindent
$^{(1)}$ 
\parbox[t]{80mm}{
t trace; - not detected}
\vspace{1mm}

\noindent
$^{(2)}$ 
\parbox[t]{80mm}{
EW of the lines of Ly$\alpha$ and CIV are measured after reconstruction
of the emission lines top.}
\vspace{1mm}

\noindent
$^{(3)}$ 
\parbox[t]{80mm}{
Galactic CII1334.5,1335.6 likely contributor to Ly$\alpha$ at $z$=0.098}

\end{table}

Several factors hamper the measurement of the absorption lines:
modest S/N, limited resolution of FOS
and uncertainties affecting the profile of the emission lines.
The absorption lines were measured assuming a plausible
reconstruction of the top
of the emission lines. Still it is not possible to obtain sufficiently
accurate Ly$\alpha$ absorption profile, and NV and CIV doublet ratios
to ascertain
whether these lines are optically thin or thick and to estimate the
covering factor (some of the galactic lines do not reach zero either).
Higher spectral resolution is needed to
elucidate these important points. With this caveat, the measures are given
in Table 4.

\subsection{The CIII$^*$1175 absorption line}

Particularly interesting is the  absorption line at 1282.5 {\AA}
(FWHM of 2{\AA} and  EW of $\approx$ 0.7 {\AA}),
which could be CIII$^*$1175 in the $z$ = 0.091 system.
The agreement in redshift is good.
This line is present in the IUE and HUT spectra of NGC 4151 (Bromage et al 1985,
Kriss et al 1992). Bromage et al (1985) argue that absorption by the
excited metastable level of CIII$^*$1175  and its strength
relative to CIV 1549
require this level to be collisionally populated in a high density medium
with N$_{e}$ = $10^{10}$ cm$^{-3}$.
In NGC 4151, the EW of CIII$^*$1175 is between 0.7 and 1.0
times the EW of CIV1548,1550 while in PG1404+226, it is
between 0.35 and 0.5 the EW of CIV1548,1550.
Note that an appealing
alternative for the identification of this line
is the {\it CIV doublet blueshifted} by 0.3 with respect to PG1404+226.
The line width, only $\approx$ 2 {\AA},
however, argues against this interpretation.
There is no other candidate for highly blueshifted absorption
lines in the spectrum.

\subsection{Origin of the UV absorption systems}

The $z$ = 0.098 system is consistent with being produced in the
halo of the host galaxy or of a nearby companion but could also
be intrinsic to the nucleus. As for the $z$ =  0.091 system
the probable presence (which needs verification) of
the  CIII$^*$1176 line suggests that
the system is intrinsic to the quasar and forms in an outflowing   wind
with a velocity of 1900 km s$^{-1}$.\\

\section{Do the UV and X-ray absorption lines come from the same absorber?}

\subsection{UV absorption lines from the Warm Absorber}

(a) The EW of the UV absorption lines expected from
the Warm Absorber have been calculated for the statistically acceptable
models of the ROSAT data.
They are given in Table 5, separately for low- and high-state
as a measure of the uncertainty arising from the continuum variability
(non simultaneous UV and X-observations, non- equilibrium of the gas,
etc. see Section 1 and Nicastro et al. 1999a).

%

\begin{figure*}
\centerline{\psfig{file=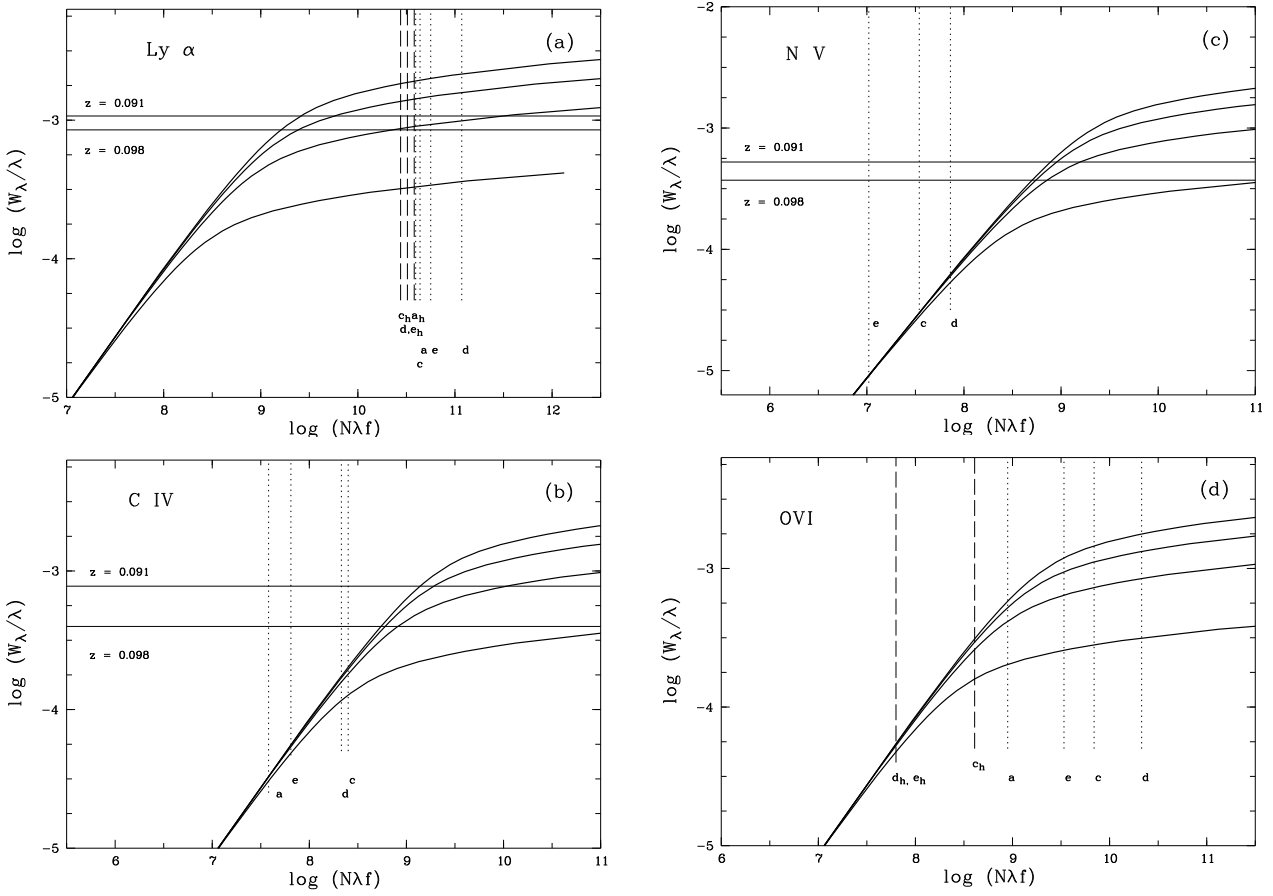,height=120mm,clip=}}

\caption{Comparison of HST UV absorption
measurements with the UV absorption in Ly$\alpha$, CIV, NV, and OVI
predicted by a warm absorber.
Thick lines: curves of growth calculated for velocity parameters
b= 20,60,100,140 km s$^{-1}$ from bottom to top (for definition of standard
quantities; see Spitzer 1978).\
Horizontal lines: Observed EW of the two UV absorption systems.
Vertical lines: predictions from the WA models (a) - (e) as defined in
Section 2.2. Predictions for the high-state data are printed as dashed lines
and the letters have an index {\it h} added, whereas predictions for the
low-state are shown as dotted lines and letters without index.
Definition of axis: W$_\lambda$ = equivalent width, $\lambda$ = wavelength,
N = column density in the relevant ion, f = oscillator strength.}
\label{}
\end{figure*}


\begin{table*}
\caption{Emission lines predicted by the warm absorber models. To judge the 
detectability of these lines, their luminosity was scaled to the one 
observed in Ly$\alpha$ and assuming 100\% covering of the warm material.
Line-ratios less than 0.01 are not listed. The emission lines given below are:
NeVIII 774, OVI 1035, FeXXI 1354, and FeXIV 5303. HS and LS refer to ROSAT
high-state and low-state, respectively.}
\label{}

      \begin{tabular}{cccccccl}
      \hline
      \noalign{\smallskip}
        state & log $U$ & log $N_{\rm w}$ &
                            \multicolumn{4}{c}{$I/I_{\rm Ly\alpha,obs}$} & model 
\\
              &         & & NeVIII & OVI &FeXXI&  FeXIV    & \\
      \noalign{\smallskip}
      \hline
      \noalign{\smallskip}
     HS/LS & 0.7/0.3 & 23.3/22.9 & 1.6 & 0.1 & 1.7 & - & (c): Ne = 
4$\times$solar abundance \\
     HS/LS & 0.7/0.5 & 23.5/23.4 & 1.3 & 0.2 & 2.5 & 0.01 & ~~~~~ O = 
0.4$\times$solar abundance \\
     HS/LS & 0.6/0.4 & 23.1/23.1 & 0.1 & 0.05& 1.2 & - & (e): additional 0.1 keV 
soft excess\\
      \noalign{\smallskip}
      \hline
  \end{tabular}

\end{table*}

We find that the CIV1550 and NV 1240 absorption lines are weak
at low state, and
negligible at high state (because C and N are highly ionized).
There is always some absorption by hydrogen due to the large column
density in H.

(b) Following Spitzer (1978), a standard curve of growth was calculated
for velocity parameters b = 20,60,100,140 km s$^{-1}$ (Figs. 8a-d).
 The predicted equivalent widths  of Ly$\alpha$,  CIV,
NV, and (OVI)  were then
compared with those derived from the analysis of the HST spectrum.

We find that at the ROSAT low-state (which is close
to the state of the source
at the time of the ASCA observations), and
for b about 60 km s$^{-1}$, there is a rough match between the
Ly$\alpha$ and CIV absorption
lines produced by the warm absorber and the observed lines at $z$ = 0.098
and $z$ = 0.090 (models c-e). The NV absorption line from the warm absorber is,
however, always weaker than observed (N is too highly ionized in the models).
Including an additional EUV bump will further increase the level of
   ionization, thus not changing the above conclusions.
   The different models differ most strongly in OVI, so this may be the
   most restrictive line, but it falls just outside the HST range.\newline
In conclusion, we find no single-phase medium which can produce both
the UV and the X-ray absorption lines.

(c) In the case of the best fit to the ASCA data i.e. the model with Fe
overabundant by a factor of $\sim$ 22 over solar, the degree of ionization
is not known but it is likely to be too high for the production of the 
CIII$^{*}$ absorption line. Thus, in this case also, the UV and X-ray 
absorbers are very probably in  different gaseous phases.

\subsection{Emission lines from the Warm Absorber}

Tab. 5 gives the intensity of the strongest lines emitted by the absorber
within the wavelength range of the HST spectra and in the optical range.
 The calculations were performed with a density of log n$_{WA}$ = 9.5.
 The line NeVIII$\lambda$774 was added to the list because of recent reports
   of its detection in high-$z$
   quasars (Hamann et al. 1997).
Table 5 is meant to provide an order
   of magnitude estimate of which lines may be important/detectable in the
future. The actual strength of the lines depends on the covering factor
of the warm absorber; total coverage was assumed for the values in Table 5.

\subsection{Relation between the X-ray and the UV absorber}

In the recent years it has been realized that gas outflow is ubiquitous in
AGN. It occurs under different gas phases: broad emission line gas (the most
highly ionized lines are the most blueshifted) and UV/optical
absorption lines (always observed at rest or blueshifted). It is likely that
the X-ray absorption features also, originate in outflowing gas at velocities
comparable with or higher than those of the UV emission/absorption gas -
this cannot presently be ascertained because of the insufficient
energy resolution of the X-ray instruments.
[This is not counting here the extraordinary
blueshift of the X-ray absorbing gas if the 1 keV features are
blueshifted OVII or OVII edges].

The question as to whether UV absorption lines and X-ray absorption lines/edges
tend to be present together or separately in AGN  has been adressed by
 Ulrich (1988) and more recently by a number of authors (e.g.
Schartel et al. 1997,  Crenshaw
1997, Mathur 1997, Shields \& Hamann 1997, Kriss et al 1996).

In some AGN,  the data appear consistent with a
single-phase, photoionized plasma producing  the UV absorption and the OVII 
and OVIII edges (NGC 3783 Shields \& Hamann 1997; 3C212, 3C351 and NGC
5548 Mathur 1997).
In contrast, in other AGN, the properties of the UV and the X-ray
absorbers imply the presence of multiphase media (NGC 3516 Kriss et al 1996;
 MCG-6-3-15 Otani et al 1996).
The present analysis indicates that  PG1404+226 is another such case of
multiphase absorbing medium
(but the non-simultaneity of the observations has to be kept in mind).

With the detection of X-ray and UV absorption in PG1404+226
the statistical association  between the presence of UV absorption lines and
X-ray absorption edges becomes stronger (Ulrich 1988, Mathur
Wilkes \& Elvis 1998). The two absorbers could be two different gaseous
phases, partaking in the same outflow but differing by their physical
conditions, velocity and radial distance to the central black hole.

\section{Conclusions}

The main results of our analysis can be summarized as follows:

\par
1) The X-ray spectrum of PG 1404+226 is variable by a factor 4 in
$\sim 3$ 10$^{4}$s
 and is characterized by a strong soft excess below 2 keV
whose luminosity ($L_{Soft} \sim 7 \times 10^{43}$ erg s$^{-1}$ in
the 0.4--2.0 keV
energy range) is about a factor
3 greater than the 2--10 keV luminosity ($L_{Hard}$).
\par
The soft excess emission can be described
with a high temperature optically thick blackbody ($kT \sim 140$ eV).
Optically thin models are ruled out combining the observed luminosity
with the dimension of the region derived from the variability timescale.
\par
2) The residuals around 1 keV can be best described by an
absorption edge at $E = 1.07 \pm 0.03$ keV, {\it not} consistent
with being caused by highly ionized oxygen at rest in the quasar frame.
A possible explanation could be either in terms of iron overabundance,
as suggested by the warm absorber fits and the extremely high EW of the 
iron K$\alpha$ line,
or by resonant absorption in a turbulent gas.
  The interpretation of a blueshifted Oxygen edge in a relativistically
  outflowing gas is less likely and not supported by the optical--UV data.
  X--ray observations of NLS1 at high spectral resolution with XMM and
  {\it Chandra} will allow to clarify the origin of the 1 keV absorption
  detected in PG 1404+226 and 3 other NLS1.

3) Two systems of absorption lines separated by $\sim 1900$ km s$^{-1}$
 are identified in the FOS/HST spectra in the
lines of Ly$\alpha$, C\,{\sc iv} and N\,{\sc v}. One system is located to the
red of the emission line peaks. Considering that in most radio quiet QSOs the
highly ionized emission lines are blueshifted (as part
of an outflow from the face of the accretion disk turned toward us) we argue
that this absorption system is nearly at rest in the AGN frame.
Its properties are consistent with this absorber being produced in the
halo of the host galaxy or a companion.
As for the system blueshifted by  $\sim 1900$ km s$^{-1}$, the very
probable detection of CIII$^{*}$$\lambda$1175 (which has been seen in
absorption only in NGC 4151 - Bromage et al. 1985, Kriss et al. 1992 )
indicates that this system is intrinsic to the quasar.

4) With the detection of X-ray and UV absorption in PG1404+226
the statistical association  between the presence of UV absorption lines and
X-ray absorption edges first suggested by Ulrich (1988) is becoming
clearer (Mathur Wilkes \& Elvis 1998). We may be seeing two different
absorbers with different physical conditions and locations but both being
parts of a grand design outflow.
The differences observed among absorption features in various AGN are
likely to result from intrinsic differences in the properties of
the gaseous outflows, from differences in the aspect angles and from 
the shape of the X--ray spectrum.

\begin{acknowledgements}

  We thank Gary Ferland for providing Cloudy.
  AC acknowledges partial support by the Italian Space Agency (ASI) under the
 contract ASI-ARS-96-70 and by the Italian Ministry for University and
  Research (MURST) under grant Cofin98-02-32.
St. K. acknowledges support from the Verbundforschung under grant
No. 50 OR 93065, and
P.C. acknowledges support from NASA through contract NAS5-26670.
\end{acknowledgements}

\end{document}